\documentstyle[aps,prl,epsfig,color,graphics]{revtex}
\newcommand{\postscript}[2]
 {\setlength{\epsfxsize}{#2\hsize} \centerline{\epsfbox{#1}}}

\begin{document}
\draft
 

\title{Search for Scalar Top and Scalar Bottom Quarks in $p\bar{p}$ collisions 
at \mbox{$\sqrt{s}=1.8$~TeV}}
\centerline{{\bf Search for Scalar Top and Scalar Bottom Quarks in $p\bar{p}$ collisions at \mbox{$\sqrt{s}=1.8$~TeV}}}

\font\eightit=cmti8
\def\r#1{\ignorespaces $^{#1}$}
\hfilneg
\begin{sloppypar}
\noindent
T.~Affolder,\r {21} H.~Akimoto,\r {43}
A.~Akopian,\r {36} M.~G.~Albrow,\r {10} P.~Amaral,\r 7 S.~R.~Amendolia,\r {32} 
D.~Amidei,\r {24} K.~Anikeev,\r {22} J.~Antos,\r 1 
G.~Apollinari,\r {36}
% ...
\end{sloppypar}
\vskip .026in
\begin{center}
(CDF Collaboration)
\end{center}

\vskip .026in
\begin{center}
\r 1  {\eightit Institute of Physics, Academia Sinica, Taipei, Taiwan 11529, 
Republic of China} \\
\r 2  {\eightit Argonne National Laboratory, Argonne, Illinois 60439} \\
\r 3  {\eightit Istituto Nazionale di Fisica Nucleare, University of Bologna,
I-40127 Bologna, Italy} \\
\r 4  {\eightit Brandeis University, Waltham, Massachusetts 02254} \\
\r 5  {\eightit University of California at Los Angeles, Los 
Angeles, California  90024} \\  
\r 6  {\eightit Instituto de Fisica de Cantabria, University of Cantabria, 
39005 Santander, Spain} \\
\r 7  {\eightit Enrico Fermi Institute, University of Chicago, Chicago, 
Illinois 60637} \\
\r 8  {\eightit Joint Institute for Nuclear Research, RU-141980 Dubna, Russia}
\\
\r 9  {\eightit Duke University, Durham, North Carolina  27708} \\
\r {10}  {\eightit Fermi National Accelerator Laboratory, Batavia, Illinois 
60510} \\
\r {11} {\eightit University of Florida, Gainesville, Florida  32611} \\
\r {12} {\eightit Laboratori Nazionali di Frascati, Istituto Nazionale
di Fisica Nucleare, I-00044 Frascati, Italy} \\
\r {13} {\eightit University of Geneva, CH-1211 Geneva 4, Switzerland} \\
\r {14} {\eightit Harvard University, Cambridge, Massachusetts 02138} \\
\r {15} {\eightit Hiroshima University, Higashi-Hiroshima 724, Japan} \\
\r {16} {\eightit University of Illinois, Urbana, Illinois 61801} \\
\r {17} {\eightit The Johns Hopkins University, Baltimore, Maryland 21218} \\
\r {18} {\eightit Institut f\"{u}r Experimentelle Kernphysik, 
Universit\"{a}t Karlsruhe, 76128 Karlsruhe, Germany} \\
\r {19} {\eightit Korean Hadron Collider Laboratory: Kyungpook National
University, Taegu 702-701; Seoul National University, Seoul 151-742; and
SungKyunKwan University, Suwon 440-746; Korea} \\
\r {20} {\eightit High Energy Accelerator Research Organization (KEK),
Tsukuba, Ibaraki 305, Japan} \\
\r {21} {\eightit Ernest Orlando Lawrence Berkeley National Laboratory, 
Berkeley, California 94720} \\
\r {22} {\eightit Massachusetts Institute of Technology, Cambridge,
Massachusetts  02139} \\   
\r {23} {\eightit Institute of Particle Physics: McGill University, Montreal 
H3A 2T8; and University of Toronto, Toronto M5S 1A7; Canada} \\
\r {24} {\eightit University of Michigan, Ann Arbor, Michigan 48109} \\
\r {25} {\eightit Michigan State University, East Lansing, Michigan  48824} \\
\r {26} {\eightit University of New Mexico, Albuquerque, New Mexico 87131} \\
\r {27} {\eightit The Ohio State University, Columbus, Ohio  43210} \\
\r {28} {\eightit Osaka City University, Osaka 588, Japan} \\
\r {29} {\eightit University of Oxford, Oxford OX1 3RH, United Kingdom} \\
\r {30} {\eightit Universita di Padova, Istituto Nazionale di Fisica 
          Nucleare, Sezione di Padova, I-35131 Padova, Italy} \\
\r {31} {\eightit University of Pennsylvania, Philadelphia, 
        Pennsylvania 19104} \\   
\r {32} {\eightit Istituto Nazionale di Fisica Nucleare, University and Scuola
               Normale Superiore of Pisa, I-56100 Pisa, Italy} \\
\r {33} {\eightit University of Pittsburgh, Pittsburgh, Pennsylvania 15260} \\
\r {34} {\eightit Purdue University, West Lafayette, Indiana 47907} \\
\r {35} {\eightit University of Rochester, Rochester, New York 14627} \\
\r {36} {\eightit Rockefeller University, New York, New York 10021} \\
\r {37} {\eightit Rutgers University, Piscataway, New Jersey 08855} \\
\r {38} {\eightit Texas A\&M University, College Station, Texas 77843} \\
\r {39} {\eightit Texas Tech University, Lubbock, Texas 79409} \\
\r {40} {\eightit Istituto Nazionale di Fisica Nucleare, University of Trieste/
Udine, Italy} \\
\r {41} {\eightit University of Tsukuba, Tsukuba, Ibaraki 305, Japan} \\
\r {42} {\eightit Tufts University, Medford, Massachusetts 02155} \\
\r {43} {\eightit Waseda University, Tokyo 169, Japan} \\
\r {44} {\eightit University of Wisconsin, Madison, Wisconsin 53706} \\
\r {45} {\eightit Yale University, New Haven, Connecticut 06520} \\
\end{center}

\begin{abstract}
We have searched for direct pair production of scalar top and scalar
bottom quarks in 88 pb$^{-1}$ of $p\overline{p}$ collisions at
$\sqrt{s}=1.8$~TeV with the CDF detector. We looked for events with a
pair of heavy flavor jets and missing energy, consistent with scalar
top quark decays to a charm quark and a neutralino, or scalar bottom
quark decays to a bottom quark and a neutralino. The numbers of events
that pass our selection for each process show no deviation from
Standard Model expectations. We compare our results to next-to-leading
order calculations for the scalar quark production cross sections to
exclude regions in scalar quark-neutralino parameter space.
\end{abstract}
\vspace{0.3cm}
\centerline{PACS numbers: 14.80.Ly, 13.85.Rm}
\vspace{0.3cm}


Supersymmetry (SUSY)~\cite{mssm} assigns to every fermionic Standard
Model (SM) particle a bosonic superpartner and to every bosonic SM
particle a fermionic superpartner.  Therefore, the SM quark helicity
states $q_L$ and $q_R$ acquire scalar partners $\tilde{q}_L$ and
$\tilde{q}_R$. SUSY models usually predict that the masses of the first
two generations of scalar quarks are approximately degenerate.  The
scalar top quark ($\tilde{t}$) mass, however, may be lower than that
of the other scalar quarks due to a substantial Yukawa coupling
resulting from the large top quark mass.  In addition, mixing between
$\tilde{t}_L$ and $\tilde{t}_R$ can cause a large splitting between
the mass eigenstates $\tilde{t}_1$ and $\tilde{t}_2$~\cite{lstop}.  If
the Yukawa coupling strength or the mixing is strong enough (or both)
then the mass eigenstate $\tilde{t}_1$ can be lighter than the top
quark.  We note that many baryogenesis models require a light stop
quark~\cite{Riotto:1999yt}.

The bottom quark mass is much smaller than the top quark mass,
therefore the effect of the Yukawa coupling on the scalar bottom quark
($\tilde{b}$) mass is small.  However, in some regions of
SUSY parameter space a large mixing between $\tilde{b}_L$
and $\tilde{b}_R$ can still occur, leading to a significant splitting
between mass eigenstates and a low mass value for the lighter mass
eigenstate ($\tilde{b}_1$)~\cite{Bartl:1994bu}.

At the Tevatron, third generation scalar quarks are expected to be
produced in pairs via $gg$ fusion and $q\overline{q}$ annihilation. In
this Letter, we describe two analyses looking for processes in a
Minimal Supersymmetric Standard Model framework: (i) a scalar top
analysis, searching for the process $p\overline{p}\rightarrow
\tilde{t}_1 \overline{\tilde{t}}_1 \rightarrow (c\tilde{\chi}_{1}^{0})\
(\overline{c} \tilde{\chi}_{1}^{0})$, and (ii) a scalar bottom
analysis, searching for the process $p\overline{p} \rightarrow
\tilde{b}_1 \overline{\tilde{b}}_1 \rightarrow (b \tilde{\chi}_{1}^{0})\
(\overline{b} \tilde{\chi}_{1}^{0})$.  We do not assume a gaugino
unification hypothesis.  We assume the lightest neutralino
$\tilde{\chi}_{1}^{0}$ is the lightest supersymmetric particle and
stable. This leads to experimental signatures with appreciable missing
transverse energy. The decay $\tilde{t}_1 \rightarrow c
\tilde{\chi}_{1}^{0}$, as in process (i), dominates via a one-loop
diagram in the absence of flavor--changing neutral currents if
$m_{\tilde{t}_1} < m_b + m_{\tilde{\chi}_{1}^{\pm}}$ and
$m_{\tilde{t}_1} < m_W + m_b + m_{\tilde{\chi}_{1}^{0}}$~\cite{lstop}.
For process (ii) we assume $m_{\tilde{b}_1} > m_b +
m_{\tilde{\chi}_{1}^{0}}$ and $m_{\tilde{b}_1} < m_b +
m_{\tilde{\chi}_{2}^{0}} $~\cite{Bartl:1994bu}.  Here,
$\tilde{\chi}_{1}^{\pm}$ and $\tilde{\chi}_{2}^{0}$ are the lightest
chargino and next-to-lightest neutralino.  Therefore, the signature of
both processes is a pair of acolinear heavy flavor jets,
$\not\!\!{E}_T$, and no leptons in the final state.

We have searched data corresponding to a total integrated luminosity
of \mbox{88.0$\pm$3.6~pb$^{-1}$} collected using the CDF detector
during the 1994-95 Tevatron run.  CDF is a general purpose detector
and is described in detail elsewhere~\cite{Abe:1988me}\@.  Here we
give a brief description of the components relevant to this analysis.
The innermost part of CDF, a four--layer silicon vertex detector
(SVX$^\prime$), allows a precise measurement of a track's impact
parameter with respect to the primary vertex in the plane transverse
to the beam direction~\cite{Cihangir:1995gx}.  A time projection
chamber determines the position of the primary vertex along the beam
direction.  The central drift chamber, located inside a 1.4--T
superconducting solenoidal magnet, measures the momenta of the charged
particles.  Outside the drift chamber there is a calorimeter, which is
organized into electromagnetic and hadronic components, with
projective towers covering the pseudo-rapidity range $|\eta|<4.2$.
The muon system is located outside the calorimeter and covers the
range $|\eta|<1$.  Events for this analysis were collected using a
trigger which required missing transverse energy $\not\!\!{E}_T >
35$~GeV.  $\not\!\!{E}_T$ is the energy imbalance in the directions
transverse to the beam direction using the raw energy deposited in
calorimeter towers with $|\eta|<3.6$.

After removing events with large $\not\!\!{E}_T$ from
accelerator--induced and cosmic ray sources, we select events with two
or three jets that have transverse energy $E_T \geq 15$~GeV and
$|\eta| \leq 2$ (hard jets) and no jets with $7 \leq E_T < 15$~GeV and
$|\eta| \leq 3.6$ (soft jets).  These requirements efficiently reject
$t\overline{t}$ events (which have more than 3 hard jets) and QCD
multijet events (which have soft jets due to gluon radiation).  Jets
are found from calorimeter information using a fixed cone
algorithm~\cite{Abe:1993rv} with a cone radius of 0.4 in
$\eta$--$\phi$ and jet energies are calculated using the raw energy
deposition in calorimeter towers.  The angle $\phi$ is the angle in
the plane normal to the beam direction. To reduce systematic effects
from the trigger, we require events to have $\not\!\!{E}_T >
40$~GeV, and to reject events with fake missing energy arising from
jet energy mismeasurements we require that the missing transverse
energy direction is neither parallel to any jet($j$) nor anti-parallel
to the leading $E_T$ jet : $\
\Delta\phi(\not\!\!{E}_T,\ j) > 45^\circ$ and
$\Delta\phi(\not\!\!{E}_T,j_1) < 165^\circ$ where the jet indices are
ordered by decreasing $E_T$.  Moreover, to reduce the QCD background, we
require the angle between the two leading jets to be
$45^\circ<\Delta\phi(j_1,j_2)<165^\circ$.  We reject events with one
or more identified electrons (muons) with $E_T\
(P_{T})>10$~GeV~(GeV$/c$).

After applying these requirements, the data sample (which we call the
pretag sample) contains 396 events.  The largest source of background
in the pretag sample is the production of $W$+jets, where the $W$
decays to a neutrino (leading to missing energy) and either an
electron or muon that is not identified or a tau which decays
hadronically.  The pretag sample also contains QCD multijet events
where the large $\not\!\!{E}_T$ is due to jet energy mismeasurement.

The SVX$^\prime$ information is used to tag heavy--flavor jets.  We
associate tracks to a jet by requiring that the track is within a cone
of 0.4 in $\eta$--$\phi$ space around the jet axis. We require tracks
to have $P_T>1.0$~GeV$/c$, positive impact parameter, and a good
SVX$^\prime$ hit pattern.  A good SVX$^\prime$ hit pattern
consists of three or four hits in the SVX$^\prime$ detector with no
hits shared by other tracks.  We take the sign of a track's impact
parameter to be the sign of the scalar product of the impact parameter
and jet $E_T$ vectors. We then define the impact parameter
significance to be the impact parameter divided by its
uncertainty. For tracks originating from the primary vertex the impact
parameter significance distribution is symmetric around zero with a
shape determined by the SVX$^\prime$ resolution, while decay products
of long lived objects tend to have large positive impact parameter
significances. We therefore use the negative impact parameter
significance distribution to define the detector resolution
function. For each track, we determine the probability that the track
comes from the primary vertex using this resolution function. We call
this probability \emph{track probability}. By construction, the
\emph{track probability} distribution is flat for tracks originating
from the primary vertex, and peaks near zero for tracks from a
secondary vertex.  We combine the track probabilities for all tracks
associated to a jet to form the \emph{jet
probability}(${\mathcal{P}}_{jet}$)~\cite{PDerwent}, the probability
that all the tracks in the jet come from the primary vertex.  The
distribution of ${\mathcal{P}}_{jet}$ is flat for jets originating
from the primary vertex by construction, while for bottom and charm
jets it peaks near zero.  ${\mathcal{P}}_{jet}$ is a continuous
variable and the ${\mathcal{P}}_{jet}$ requirement is easily optimized
for the scalar top and scalar bottom searches separately.  This
motivates the choice of the ${\mathcal{P}}_{jet}$ tagging algorithm
over other tagging algorithms developed at CDF~\cite{Abe:1995hr}.

\begin{figure}[p]
  \begin{center} \postscript{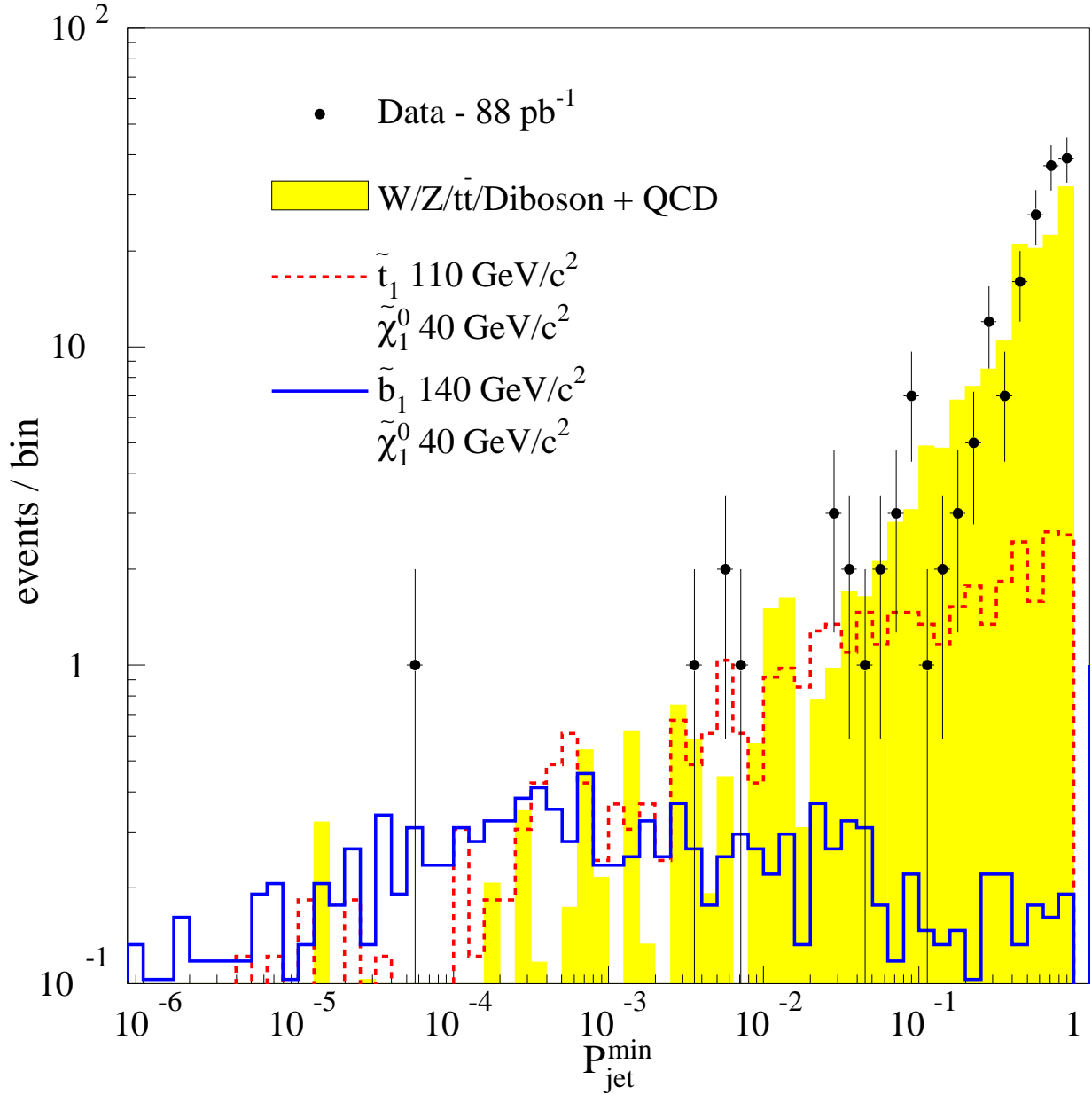}{1.}  \end{center}
  \caption{The distribution of ${\mathcal{P}}_{jet}^{min}$ - the lowest
  value of ${\mathcal{P}}_{jet}$ for all taggable jets in an event. A
  requirement of ${\mathcal{P}}_{jet}^{min} \leq 0.05\ (0.01)$ is
  applied to select charm (bottom) jets. Points are data, the shaded
  histogram is the sum of the predicted backgrounds, the solid line is
  the predicted signal for \mbox{$m_{\tilde{t}_1}=110$~GeV$/c^2$},
  \mbox{$m_{\tilde{\chi}_{1}^{0}}=40$~GeV$/c^2$}, and the dashed line
  is the predicted signal for \mbox{$m_{\tilde{b}_1}=140$~GeV$/c^2$},
  \mbox{$m_{\tilde{\chi}_{1}^{0}}=40$~GeV$/c^2$}.  The background and
  signal are normalized to 88 pb$^{-1}$.}  \label{fig:jpb}
\end{figure}

We select events for the scalar top analysis by requiring the
event to have at least one taggable jet with a ${\mathcal{P}}_{jet}
\leq 0.05$. A taggable jet has at least two SVX$^\prime$
tracks as defined above.  The distribution of the minimum jet
probability (${\mathcal{P}}_{jet}^{min}$) of the taggable jets in the
pretag sample is shown in Figure~\ref{fig:jpb}.  The expected
background and scalar top/bottom signal distributions are overlaid.
This requirement, chosen to optimize the expected signal significance,
rejects approxiamtely 97\% of the background while its efficiency for
the signal is 25\%.  For the scalar bottom analysis the expected
signal significance is optimized by requiring that the event has at
least one taggable jet with a ${\mathcal{P}}_{jet}^{min} \leq 0.01$.
This requirement rejects approximately 99\% of the background while
retaining 45\% of the scalar bottom signal.

Backgrounds (other than QCD multijet events) and the expected signal
are estimated using a number of Monte Carlo programs followed by a
full CDF detector simulation.  Single vector boson production and
decay is simulated using a tree--level calculation as implemented in
the VECBOS~\cite{Berends:1991ax} package, with HERWIG~\cite{Marchesini:1996vc}
routines used for subsequent parton hadronization.  Vector boson pair production
and decay is implemented in ISAJET~\cite{Baer:1993ae}.  Top pair
production and decay is simulated using HERWIG.  Signal events are modeled using the
PYTHIA~\cite{Sjostrand:1994yb} generator.  The PYTHIA Monte Carlo
generator includes production and decay of SUSY
particles~\cite{Mrenna:1997hu}.  The next--to--leading order (NLO)
cross section for the scalar quark production is calculated using the
PROSPINO~\cite{Beenakker:1997ut} program with CTEQ3M parton
distribution functions~\cite{Huston:1995vb}.  Simulated events are
analyzed using the same procedure as the selected data sample.  We
check the single vector boson normalization with data by reversing the
lepton veto requirement in the pretag sample.

We estimate the number of QCD multijet events in the tagged samples
using a combination of Monte Carlo and data samples.  We attribute the
excess of data events above electroweak sources in the pretag data
sample to QCD multijet sources.  The total expected electroweak
background in the pretag sample is $270.1\pm76.2$ which gives us an
estimate of $125.9\pm83.4$ expected QCD multijet events in the pretag
sample.  We then apply a ${\mathcal{P}}_{jet}$ mistag matrix to this
excess to estimate the QCD multijet background after tagging.  The
${\mathcal{P}}_{jet}$ mistag matrix, which parameterizes the
probability that a jet has ${\mathcal{P}}_{jet} \leq 0.05$ as a
function of jet $E_T$ and the number of SVX$^\prime$ tracks, is derived from
data and verified in several control data samples.

The systematic uncertainties on the expected number of signal events
apply for both $\tilde{t}_1$ and $\tilde{b}_1$.  The NLO cross section
for third generation scalar quarks depends weakly on other masses and
parameters ($\sim 1\%$)~\cite{Beenakker:1997ut}.  The dominant NLO
uncertainties are due to the choice of QCD renormalization scale
($\mu$) and the choice of parton distribution function.  The
theoretical uncertainty on the NLO scalar quark production cross
section is a function of the scalar quark mass and ranges from $11 \%$
to $22 \%$ for the mass range 30 GeV/$c^2$ to 150 GeV/$c^2$.  Gluon
radiation from the initial state (ISR) or final state (FSR) partons is
the largest source of systematic uncertainty.  We determine its effect
on our acceptance by turning off ISR or FSR in the signal Monte Carlo
and comparing the efficiency with the default Monte Carlo which has
ISR and FSR turned on.  The combined ISR/FSR systematic uncertainty is
$23\%$.  We determine the jet energy systematic uncertainty, which is
10\%, by varying the jet energies by $\pm 5\%$.  The trigger
efficiency systematic uncertainty, which is 10\%, is determined by
varying the trigger efficiency curve (which is derived from data) by
$\pm 1 \sigma$ of its fitted values.  The detection efficiency
estimates are derived from Monte Carlo that has exactly one primary
vertex.  The dominant effect of multiple primary vertices is to reduce
the efficiency for a requirement of no extra jets with $E_T
\geq 7$ GeV and $|\eta| \leq 3.6$.  We account for the loss in
efficiency due to the extra jet veto by combining the Monte Carlo with
a minimum--bias data sample (consistent with the number of primary
vertices found during the 1994-95 Tevatron run), measuring the
relative loss in efficiency of the ``no extra jet'' requirement, and
degrading the signal efficiency by this factor. The efficiency scale
factor due to multiple primary vertices is $0.93\pm0.03$.  We use data
samples enriched in charm(bottom) jets to determine the systematic
uncertainty on the charm (bottom) tagging efficiency.  The systematic
uncertainty is 10\% for both charm and bottom tagging.  Including the
systematic uncertainties due to the integrated luminosity measurement
(4.1\%) and finite Monte Carlo statistics (5--15\%), the total
systematic varies from 31\% to 36\% as a function of the squark mass.

\begin{figure}[p]
  \begin{center} \postscript{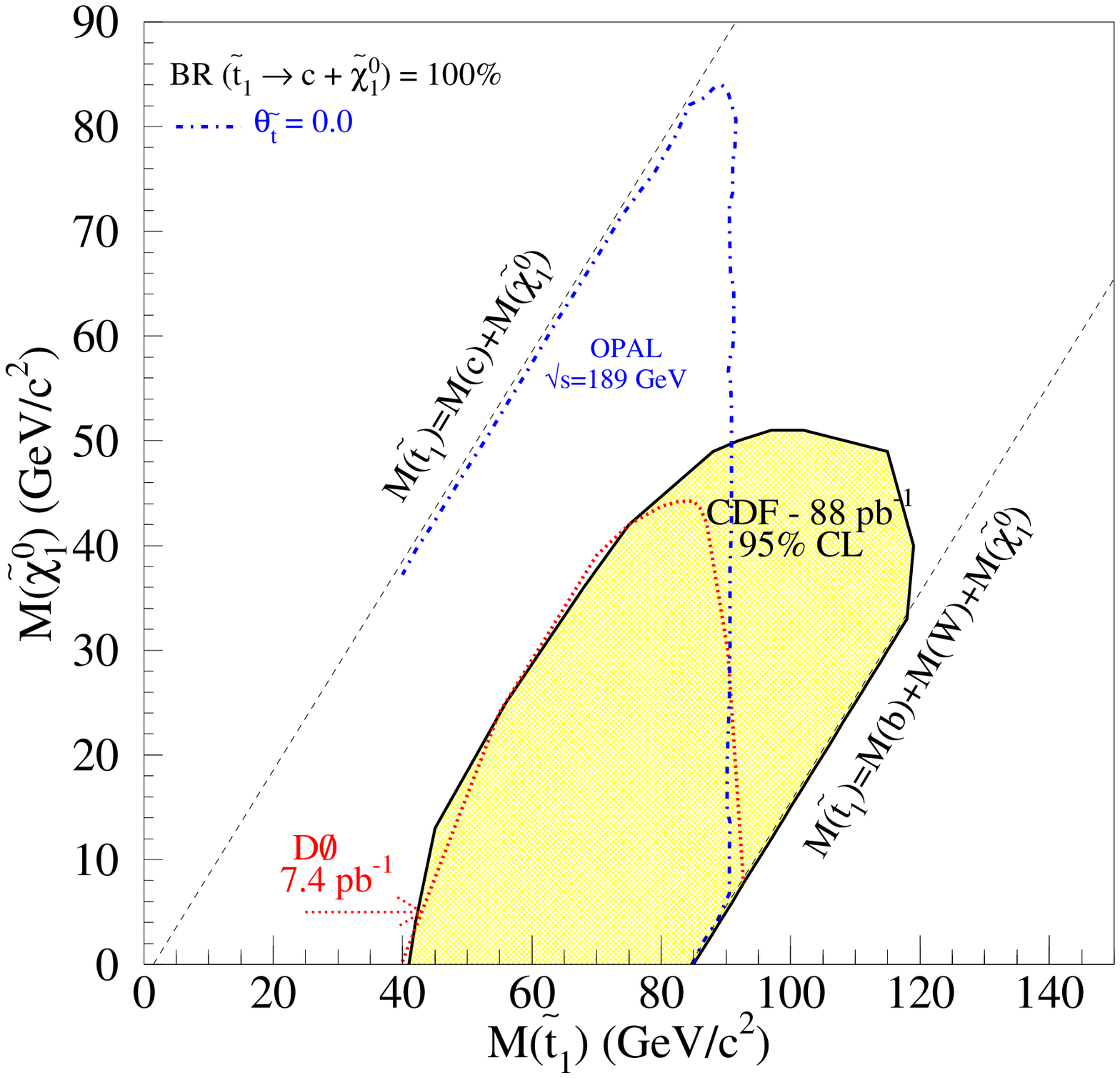}{1.}  \end{center}
  \caption{95\% CL exclusion region (shaded region) in
  $m_{\tilde{\chi}_1^0}$--$m_{\tilde{t}_1}$ plane for $\tilde{t}_1
  \longrightarrow c \tilde{\chi}_1^0$.  Also shown are results from
  D\O~\protect\cite{Abachi:1996jp} and
  OPAL~\protect\cite{Abbiendi:1999}.  $\theta_{\tilde{t}}$ parameterizes the
  mixing of the left/right scalar top gauge eigenstates to form the
  light/heavy mass eigenstates.  Note that the results for both D\O\
and CDF are independent of $\theta_{\tilde{t}}$.}
\label{fig:limit_stop}
\end{figure}

In the scalar top analysis we observe 11 events, which is consistent
with $14.5\pm4.2$ events expected from SM processes (see
Table~\ref{tbl:nolep_ptag}).  We interpret the null result in the
scalar top search as an excluded region in
\mbox{$m_{\tilde{\chi}_{1}^{0}}$--$m_{\tilde{t}_1}$}\ parameter space
using a background subtraction method~\cite{Caso:1998tx}.  The 95\%
C.L. excluded region is shown in Figure~\ref{fig:limit_stop}.  The
maximum $m_{\tilde{t}_1}$ excluded is 119~GeV$/c^2$ for
\mbox{$m_{\tilde{\chi}_{1}^{0}}$ = 40~GeV$/c^2$}.  The maximum
excluded value of the neutralino mass is 51~GeV$/c^2$ which
corresponds to the scalar top mass of 102~GeV$/c^2$. The reach in
$m_{\tilde{t}_1}$ is limited by the statistics, while the
gap between the kinematic limit and the excluded region is mostly
determined by the $\not\!\!{E}_T$ cut which is effectively fixed by
the $\not\!\!{E}_T$ trigger threshold.  Also shown in
Figure~\ref{fig:limit_stop} are the results from the D\O\ experiment,
based on 7.4~pb$^{-1}$~\cite{Abachi:1996jp}, and from the OPAL
experiment for \mbox{$\sqrt{s}=189$~GeV} at
LEP~\cite{Abbiendi:1999}.

\begin{figure}[p]
  \begin{center} \postscript{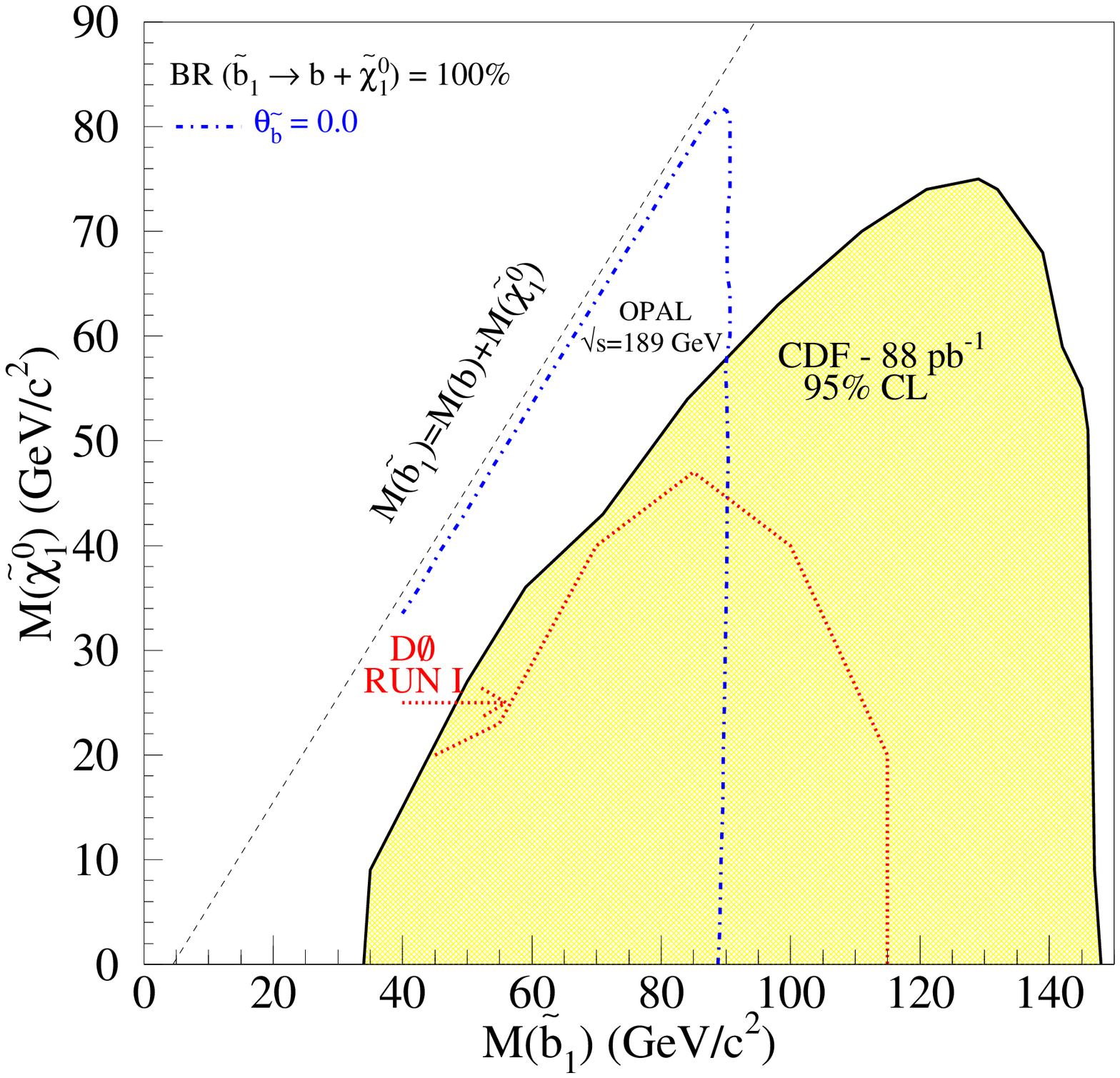}{1.}  \end{center}
  \caption{95\% CL exclusion region (shaded region) in
  $m_{\tilde{\chi}_1^0}$--$m_{\tilde{b}_1}$ plane for $\tilde{b}_1
  \longrightarrow b \tilde{\chi}_1^0$.  Also shown are the results
  from D\O~\protect\cite{Abbott:1999wt} and
  OPAL~\protect\cite{Abbiendi:1999}.  $\theta_{\tilde{b}}$ parameterizes the
  mixing of the left/right scalar bottom gauge eigenstates to form the
  light/heavy mass eigenstates.  Note that the results for both D\O\
and CDF are independent of $\theta_{\tilde{b}}$.}
\label{fig:limit_sbot}
\end{figure}

In the scalar bottom analysis five events are observed with an expected
background of $5.8\pm1.8$ (see
Table~\ref{tbl:nolep_ptag})\@. Similarly, we interpret the null result
as an excluded region in $m_{\tilde{\chi}_{1}^{0}}$--$m_{\tilde{b}_1}$ parameter space as shown
in Figure~\ref{fig:limit_sbot}.  For \mbox{$m_{\tilde{\chi}_{1}^{0}}$
= 40 GeV$/c^2$} the maximum $m_{\tilde{b}_1}$ excluded is
146~GeV$/c^2$ .  Also plotted are the latest results from
D\O~\cite{Abbott:1999wt} and OPAL~\cite{Abbiendi:1999}.

In summary, we have performed a search for $\tilde{t}_1/\tilde{b}_1$
in $p\overline{p}$ collisions at $\sqrt{s} = 1.8$ TeV using 88
pb$^{-1}$ of data.  We looked for events with significant missing
energy, no high $P_T$ lepton(s), and two or three jets.  We required
that at least one jet is consistent with originating from a heavy
flavor jet using a technique called jet probability.  After applying
all selection criteria, we observed no excess of events above Standard
Model predictions, and we set 95\% C.L. exclusion regions in the
$m_{\chi_1^0}$--$m_{\tilde{q}_1}$ plane.

We thank the Fermilab staff and the technical staffs of the
participating institutions for their vital contributions.  This work
was supported by the U.S. Department of Energy and National Science
Foundation; the Italian Istituto Nazionale di Fisica Nucleare; the
Ministry of Education, Science and Culture of Japan; the Natural
Sciences and Engineering Research Council of Canada; the National
Science Council of the Republic of China; the A. P. Sloan Foundation;
and the Swiss National Science Foundation.


\newpage
\begin{table}
 \begin{tabular}{|ldd|}
 \multicolumn{1}{|c}{Sample} & $N_{exp}$ (${\mathcal{P}}_{jet}^{min}
\leq$ 0.05) & $N_{exp}$ (${\mathcal{P}}_{jet}^{min} \leq$ 0.01) \\ \hline
 $W^{\pm}(\rightarrow e^{\pm}\nu_{e})+\geq 2$ jets & 0.3$\pm$0.3$\pm$0.1 & - \\
 $W^{\pm}(\rightarrow \mu^{\pm}\nu_{\mu})+\geq 2$ jets &
 0.9$\pm$0.5$\pm$0.3 & - \\
 $W^{\pm}(\rightarrow \tau^{\pm}\nu_{\tau})+\geq 1$ jets &
7.6$\pm$1.6$\pm$2.2 & 3.0$\pm$1.0$\pm$0.9\\ 
 $Z^{0}(\rightarrow \nu\overline{\nu})+\geq 2$ jets & 1.2$\pm$0.4$\pm$0.4 &
 0.8$\pm$0.3$\pm$0.2 \\ 
 $t\overline{t}$ & 0.7$\pm$0.2$\pm$0.4 & 0.5$\pm$0.2$\pm$0.2 \\ 
 $Diboson\ (WW,WZ,ZZ)$ & 0.4$\pm$0.1$\pm$0.1 & 0.2$\pm$0.1$\pm$0.1 \\ \hline
 Total $W/Z/t\overline{t}/Diboson$ & 11.1$\pm$1.8$\pm$3.3 &
4.5$\pm$1.1$\pm$1.2 \\ \hline 
 Total $QCD$ & 3.4$\pm$1.7 & 1.3$\pm$0.7 \\ \hline
 Total Expected from SM& 14.5$\pm$4.2 & 5.8$\pm$1.8 \\ \hline
 Total Observed & $11$& $5$ \\
 \end{tabular}
\caption {The number of observed data and expected background events.
For $W/Z/t\overline{t}/Diboson$, the first uncertainty is statistical,
the second is systematic.  For $QCD$ and Total Expected from SM, the
uncertainty is statistical plus systematic.}

\label{tbl:nolep_ptag}
\end{table}

\begin{references}

\bibitem{mssm} For reviews of SUSY and the MSSM, see
H.P.~Nilles, 
Phys. Rept. {\bf 110}, 1 (1984); H.E.~Haber and G.L.~Kane, 
Phys. Rep. {\bf 117}, 75 (1985); S.P.~Martin, ``A Supersymmetry
Primer", hep-ph/9709356; G.L.~Kane, ``Perspectives on Supersymmetry",
{\it Singapore, Singapore: World Scientific (1998) 479 p}.


\bibitem{lstop}
See, for example, K.~Hikasa and M.~Kobayashi, Phys. Rev. {\bf D36},
724 (1987); H.~Baer {\it et al.}, Phys. Rev. {\bf D44}, 725 (1991);
H.~Baer, J.~Sender and X.~Tata, Phys. Rev. {\bf D50}, 4517 (1994).

\bibitem{Riotto:1999yt}
A.~Riotto and M.~Trodden, ``Recent Progress in Baryogenesis'',
\emph{to appear in the Annual Review of Nuclear and Particle Science
Vol.49.}

\bibitem{Bartl:1994bu}
A.~Bartl, W.~Majerotto and W.~Porod, Z. Phys. {\bf C64}, 499 (1994).

\bibitem{Abe:1988me}
CDF Collaboration, F.~Abe {\it et al.}, Nucl. Instr. Meth. {\bf A271}, 387 (1988); F. Abe {\it et
al.}, Phys. Rev. {\bf D50}, 2966 (1994).

\bibitem{Cihangir:1995gx}
CDF Collaboration, S.~Cihangir {\it et al.}, Nucl. Instrum. Meth. {\bf A360}, 137 (1995).

\bibitem{Abe:1993rv}
CDF Collaboration, F.~Abe {\it et al.}, Phys. Rev. {\bf D47}, 4857
(1993).

\bibitem{PDerwent}
CDF Collaboration, F.~Abe {\it et al.}, Phys. Rev. {\bf D53}, 1051 (1996).

\bibitem{Abe:1995hr}
CDF Collaboration, F.~Abe {\it et al.}, Phys. Rev. Lett. {\bf 74}, 2626 (1995).

\bibitem{Berends:1991ax}
F.A.~Berends, H.~Kuijf, B.~Tausk and W.T.~Giele, Nucl. Phys. {\bf B357}, 32
(1991); W.T.~Giele, E.W.~Glover and D.A.~Kosower, Nucl. Phys. {\bf B403},
633 (1993).

\bibitem{Marchesini:1996vc}G.~Marchesini, B.R.~Webber, G.~Abbiendi, I.G.~Knowles,
M.H.~Seymour and L.~Stanco, Comput. Phys. Commun. {\bf67} 465 (1992).

\bibitem{Baer:1993ae}
H.~Baer, F.E.~Paige, S.D.~Protopopescu and X.~Tata,
``Simulating Supersymmetry with ISAJET 7.0 / ISASUSY 1.0,"
hep-ph/9305342.

\bibitem{Sjostrand:1994yb}
T.~Sjostrand, ``High-energy Physics Event Generation with PYTHIA 5.7
and JETSET 7.4," Comput. Phys. Commun. {\bf 82}, 74 (1994).

\bibitem{Mrenna:1997hu}
S.~Mrenna, Comput. Phys. Commun. {\bf 101}, 232 (1997).

\bibitem{Beenakker:1997ut}
W.~Beenakker {\it et al.}, 
Nucl. Phys. {\bf B515}, 3 (1998).

\bibitem{Huston:1995vb}
CTEQ Collaboration, J.~Huston {\it et al.}, Phys. Rev. {\bf D51}, 6139 (1995).

\bibitem{Caso:1998tx} C.~Caso {\it et al.}, Eur. Phys. J. {\bf C3}, 1 (1998).

\bibitem{Abachi:1996jp}
D\O\ Collaboration, S.~Abachi {\it et al.}, Phys. Rev. Lett. {\bf 76}, 2222 (1996).

\bibitem{Abbiendi:1999}
OPAL Collaboration, G.~Abbiendi {\it et al.}, Phys. Lett., {\bf B456}, 95 (1999).

\bibitem{Abbott:1999wt}
D\O\ Collaboration, B.~Abbott {\it et al.}, Phys. Rev. {\bf D60}, 031101 (1999).


\end{references}
\end{document}